\documentclass[twocolumn,showpacs,preprintnumbers,amsmath,amssymb]{revtex4}

\usepackage{graphicx}
\usepackage{dcolumn}
\usepackage{bm}

\begin{document}


\title{Random Holographic ``Large Worlds" With Emergent Dimensions}


\author{Carlo A. Trugenberger}
\email{ca.trugenberger@bluewin.ch}
\affiliation{%
SwissScientific, chemin Diodati 10, CH-1223 Cologny, Switzerland
}%

\date{\today}

\begin{abstract}
I propose a random network model governed by a Gaussian weight corresponding to Ising link antiferromagnetism as a model for emergent quantum space-time. In this model, discrete space is fundamental, not a regularization, its spectral dimension $d_s$ is not a model input but is, rather, completely determined by the antiferromagnetic coupling constant. Perturbative terms suppressing triangles and favouring squares lead to locally Euclidean ground states that are Ricci flat "large worlds" with power-law extension. I then consider the quenched graphs of lowest energy for $d_s=2$ and $d_s=3$ and I show how quenching leads to the spontaneous emergence of embedding spaces of Hausdorff dimension $d_H=4$ and $d_H=5$, respectively. One of the additional, spontaneous dimensions can be interpreted as time, causality being an emergent property that arises in the large $N$ limit (with $N$ the number of vertices). For $d_s=2$, the quenched graphs constitute a discrete version of a 5D-space-filling surface with a number of fundamental degrees of freedom scaling like $N^{2/5}$, a graph version of the holographic principle. These holographic degrees of freedom can be identified with the squares of the quenched graphs, which, being these triangle-free, are the fundamental area (or loop) quanta.  
\end{abstract}
\pacs{04.60.-m; 02.10.Ox}
\maketitle

\section{Introduction}
It is generally believed that most dynamically generated networks are typically "small worlds", with average distances (or diameters) scaling logarithmically with the volume, the total number N of vertices. This is definitely so for random graphs \cite{graphrev} which have been recently the subject of intensive study as models of interrelations in a wide variety of complex phenomena. Random graphs are often contrasted to regular lattices, which are rigid structures unable to accommodate the complex relations typical of self-organizing systems. Recently, however, there has been an increasing interest in applying the physics of graphs and networks to the problem of quantum gravity \cite{bia1}. 

Contrary to the traditional approach based on triangulations \cite{triang}, in this new approach based on networks, space-time is considered as an emergent property of a large graph: the correct space-time is expected to self-organize according to the rules of the network and is thus described by emergent topologies and geometries. A new field is thus emerging, in which complex network systems are used to model topology and geometry. 

First attempts in this direction go back to string nets \cite{wen} and the models of quantum networks introduced in \cite{graphity} and recently revisited in \cite{bia2}. It has also been shown that geometry can emerge on the boundary of random tensor networks \cite{sasakura}. In two recent publications \cite{tru1, tru2} I have proposed an alternative route based on ideas of statistical field theory and inspired by Kazakov's famed random lattice Ising model \cite{kazakov}. The idea is to couple the classical ferromagnetic Ising model to an antiferromagnetic Ising model for the links of the network on which the Ising model is defined, the binary values of the degrees of freedom denoting presence/absence of space-time and of links between the space-time points. This is a generalization of Kazakov's random lattice Ising model \cite{kazakov} in which the network topology is not restricted to be two-dimensional and the network structure is drawn according to a Gaussian distribution rather than a uniform one. It is also related to the coloured random graphs introduced in \cite{gorski}, although here graphs assemble due to a minimum energy principle that suppresses triangles and the whole dynamics will depend, rather on squares. 

The competition between vertex ferromagnetism and link antiferromagnetism creates frustration and this leads to the formation of a ground state network with very interesting properties, the connectivity $k$, and thus, the geometric dimensionality being determined uniquely by the coupling constant $J$. In the ferromagnetic phase for the vertices, which is interpreted as a fully formed space-time, this frustration becomes simply an external magnetic field for the link antiferromagnet and the model simplifies considerably. In \cite{tru2} I have solved this model on a bipartite lattice in the mean field approximation.

There is a line of second-order phase transitions $\left( J_{\rm cr} ,T_{\rm cr} \right)$. Above the critical line, the network is purely ferromagnetic, with all links of the lattice having the same probability of existing. Below the critical line, the network starts to develop antiferromagnetism, in the sense that, at each vertex some of the links decouple and develop a high probability of being absent. At zero temperature, and below a critical value of the coupling $J_{\rm cr} (T=0)$ the network consists typically of a space-filling $d$-dimensional discrete hypersurface embedded in the original $D$-dimensional lattice, a higher-dimensional generalization of a surface-filling Peano or Hilbert curve \cite{hilbert}.

In \cite{tru1} I have provided first numerical evidence that this picture is realized also in the full model, with no restrictions to a fixed background $D$-dimensional bipartite lattice and without the mean field approximation. Indeed, the numerical evidence shows that, for connectivities below the critical value 8, the ground state graphs have a spectral dimension $d_s$ that continues to be determined by the coupling constant, while the Hausdorff dimension decouples so that $d_H>d_s$, which is exactly what would happen if the ground state would consist of a $d_H$-dimensional-space-filling lower-dimensional graph. 

The ground state graphs of the model, relevant for the zero temperature physics are always regular graphs, i.e. all vertices have the same connectivity. It is known that random regular graphs, sampled from a uniform distribution, are so called "small worlds" i.e. have a diameter and an average distance that scale logarithmically with the total number N of vertices \cite{rrg}. This behaviour is unsuitable for a discrete approximation of a finite-dimensional universe and seems to be incompatible with the numerical results in \cite{tru1}, which, however, were restricted to small samples where it is difficult to distinguish numerically between a logarithmic behaviour and a small power law. In \cite{tru1}, the highly degenerate ground states corresponding to regular graphs were sampled by a Glauber dynamics starting from a random initial configuration of very high energy. This is 
what goes under the name of quench in statistical mechanics \cite{quench} and represents a sudden, fast cooling of the system. It is not clear that this sampling is equivalent to the uniform sampling of random regular graphs and this may explain the discrepancy observed. 

In this paper I will concentrate on the zero-temperature physics of the model, i.e. on its classical limit. In particular I will suggest to add a perturbative term to the Hamiltonian which has important consequences in this limit while being irrelevant for higher temperatures. With this add-on, the discrepancy described above can be clearly resolved in favour of a power law corresponding to a finite Hausdorff dimension, a logarithmic scaling of average distances being ruled out. These ground state graphs are thus "large worlds" although they are random structures, sampled typically from Gaussian distributions. For particular values of the coupling constant, moreover, one can show that the local geometric structure of the ground state graphs is locally Euclidean, analogous to that of a regular square or cubic lattice. This has two immediate consequences, there are no triangles and there is a perfect matching between the unit spheres of neighbouring vertices. These properties imply that both recently developed concepts of combinatorial Ricci curvature for graphs, the Ollivier curvature \cite{olli1, olli2, liu} and the Knill curvature \cite{knill} vanish and so does then also the Ricci curvature obtained in the continuum limit. These "large world" ground state graphs are thus Ricci flat. 

The energy value determines only the local geometric structure of the ground state. This still leaves a large ground state degeneracy among all graphs with the same local geometry but different global properties. In particular, the topology, describing if and how the graph is globally curled up in a possible embedding dimension is not determined by the Hamiltonian. The same regular graph of connectivity 4 drawn on a flat sheet of paper or on a sheet of paper that is subsequently crumpled up in a ball have the same energy, but very different topologies, characterized by a different Hausdorff dimension (two in the former case, three in the latter). The topological structure depends thus not only on the Hamiltonian but also on how the fluctuations within the degenerate ground state manifold are sampled. As in \cite{tru1} I will consider quenching the graphs from very high energy random configurations by a Glauber dynamics \cite{quench}. It is well known that quenching can lead to complex behaviour when various phases can coexist at low temperatures. Here I will show how it leads, for coupling constants implying 3d ground state graphs, to the spontaneous emergence of a 4D embedding space, implying a Hausdorff dimension $d_H=4$ for quenched graphs of sufficiently low energy. This implies the emergence of one "spontaneous" dimension, which I propose to identify with time. This identification is supported by the application of very recent mathematical results on the number of loops on generic regular graphs \cite{loopreg1, loopreg2}. Applying these results I will show that no closed loops with a number of steps in the emergent time direction below a critical value survive in the continuum limit and I will identify these open paths as "causal" trajectories. I propose thus that causality is not a fundamental property of space-time but, rather, it is only an emergent feature at large scales. 

Finally, my results offer a new interpretation of the holographic principle \cite{holo}, the by now well-established notion that in a proper quantum gravity theory the number of fundamental degrees of freedom is related to the area of surfaces in space-time rather than the volume. The holographic principle is mostly discussed and explained in terms of black hole physics or universes with boundaries, where there is a naturally defined surface to consider, either the Schwarzschild radius of the black hole or the boundary of the universe. In the present model, however it arises completely naturally for coupling constants implying 2d ground states, for which there are three emergent dimensions, implying $d_H=5$. The number of degrees of freedom in the 5D space scales as the area of the space-filling holographic surface, a graph version of the holographic principle. As I will show, these holographic degrees of freedom are the fundamental areas/loops on the quenched holographic screen.

\section{Dynamical regular graphs}
The model is formulated in terms of $N$ bits $s_i = \pm 1$, for $i=1\dots N$ and $N(N-1)/2$ bits $w_{ij}=w_{ji}=0,1$, for
$i,j=1\dots N$. A value $s_i=+1$ denotes the existence of space-time, while $s_i=-1$ indicates the absence of space time (or the presence of anti-space-time). A value $w_{ij}=1$ denotes a connection between bits $s_i$ and $s_j$, a value $w_{ij}=0$ indicates that $s_i$ and $s_j$ are not connected. These link variables are symmetric, $w_{ij} =w_{ji}$ and vanish on the diagonal, $w_{ii} = 0$.

The Hamiltonian for the coupled Ising-network system is (I use natural units $c=1$, $\hbar =1$)
\begin{equation}
H_0 =J_s \left[  {J\over 2} \ \sum_{i\ne j} \sum_{k\ne i \atop k\ne j} w_{ik}w_{kj} - {1\over 2} \sum_{i\ne j} s_i w_{ij} s_j \right] \ , 
\label{twob}
\end{equation}
where the Ising coupling $J_s$ represents the Planck energy scale $\hbar /t_P$ ($t_P$ being the Planck time) and $J$ is the unique dimensionless coupling. For simplicity, from now on I will also set $J_s=1$. 

The second term in the Hamiltonian is the standard ferromagnetic Ising model. The first term, instead represents a nearest-neighbours antiferromagnetic Ising model for the links, with "nearest-neighbours" meant in the sense that only links sharing a common vertex are coupled. 
In absence of the antiferromagnetic link term, and with the links $w_{ij}$ uniformly drawn from random adjacency matrices of degree 4 , the model would be Kazakov's random lattice Ising model in two dimensions \cite{kazakov}.  The generalizations with respect to Kazakov's model, thus consist in dropping the restriction to degree 4 and drawing the random adjacency matrices from a Gaussian distribution.

The competition between the vertex ferromagnetic coupling and the link antiferromagentic one generates frustration in the model. Indeed the vertex ferromagnetic coupling favours the creation of many links (positive values of $w_{ij}$) in a vertex-aligned state, corresponding to a fully formed space-time: space-time points have a tendency to link together. On the other side, due to the antiferromagnetic link coupling, creating many links costs energy. The compromise is to create a uniform, optimal number of links per vertex depending on the coupling constant $J$, generating thus a regular graph. 

To show this, let me consider a $k$-regular graph, i.e. a configuration with $s_i = +1$, $\forall i$ and such that each vertex has exactly $k$ incident edges (degree $k$ in graph parlance \cite{graphrev}). I will now show that, for a particular range of the coupling constant $J$, such a graph is a local minimum of the energy (\ref{twob}). To this end, let me suppose that such a graph is given and let me first change one single vertex spin from $s_i=+1$ to $s_i=-1$. The corresponding energy change $\Delta E_i = k > 0$ is positive. Changing vertex spins, thus costs energy. Let me now try to eliminate an existing connection, i.e. changing $w_{ij} = 1$ to $w_{ij} = 0$. In this case there are contributions from both terms in the energy function. Before the elimination, the existing connection contributed $E_{ij} = J(k-1)-(1/2)$. After the elimination, of course the contribution of this connection vanishes. The energy change due to the elimination of a connection $w_{ij}$ is thus $\Delta_{\rm elim} E_{ij} = (1/2)- J(k-1)$. Let me finally consider adding a previously non-existent connection $w_{ij}$. In this case it is the energy contribution before the addition that vanishes, while, after the addition I have an additional energy $E_{ij}= Jk -(1/2)$. The energy change for adding a connection is thus $\Delta_{\rm add} E_{ij} = Jk-(1/2)$. Requiring that both eliminating and adding a connection costs energy we obtain the stability condition
\begin{equation}
{1\over 2J} < k < 1+{1\over 2J} \ .
\label{twoc}
\end{equation}
To proceed, let me compute the total energy of a $k$-regular graph on $N$ vertices. This is easily obtained as
\begin{equation}
E_{N, k} = N \left( {J\over 2} k (k-1) -{k\over 2} \right) \ .
\label{twod}
\end{equation}
This expression is minimized when the vertex degree takes the value $k=1/2 + (1/2J)$ and this value of $k$ satisfies the stability condition (\ref{twoc}). Defining 
\begin{equation}
J = {1\over 4d-1} \ ,
\label{twoe}
\end{equation}
I have obtained the result that for any choice of half-integer $d$ a $2d$-regular graph is a local minimum of the energy and that this minimum is the one of lowest energy among all possible regular graphs. Of course, this is not yet a complete proof that $2d$-regular graphs are the true global minima of the energy functional (\ref{twob}). There could be other, non-regular graphs, with even lower energies. That this is not so can be confirmed by numerical computations \cite{tru1}. 

Actually, subsequent and more detailed numerical analyses \cite{fabio} have revealed an even more interesting structure for generic $d$. The ground state graphs are, indeed, regular graphs for all, generic real values of $d$, and thus $J$. The integer connectivity is given by $k = {\rm Round} \left( 1/2 + (1/2J) \right)$, i.e. by the nearest integer approximation to the real value $1/2 + (1/2J)$. As I will show, the function $D_g={\rm ceiling} (k/2)$ can thus be considered as the geometric dimension of space-time. As anticipated, in this model the dimension of space-time is dynamical, rather than a fixed input parameter. Indeed, it is nothing else than the unique dimensionless coupling constant. 

\section{Dynamical locally Euclidean graphs}
Random regular graphs \cite{rrg} are regular graphs drawn from a uniform probability distribution. It is known that random regular graphs are "small worlds", i.e. their diameter and average distances on the graphs scale logarithmically with the number N of vertices (the volume). This behaviour is unsuitable to model space-times. It is not clear that the Glauber dynamics sampling using in \cite{tru1} to obtain the ground state graphs of the model (\ref{twob}) is indeed equivalent to a uniform distribution and does not favour, instead, a particular subset of regular graphs. In any case, I propose here a small modification to the Hamiltonian (\ref{twob}) that explicitly selects only a particular subset of regular graphs. 

Loosely speaking, random regular graphs have locally a tree structure \cite{rrg} and thus the lack of links used to form short cycles leaves lots of links available to form "shortcuts" among otherwise "distant" parts of the graph, causing the logarithmic scaling behaviour. I will thus add a small perturbation $H_1$ to the Hamiltonian (\ref{twob}) that favours the formation of squares, while suppressing the triangles associated typically with curvature: 
\begin{eqnarray}
H &&= H_0 + H_1 \ ,
\nonumber \\
H_1 &&= {\lambda_3 \over 3} {\rm Tr} \left( w^3 \right) -{\lambda_4 \over 4} \ {\rm Tr} \left( w^4\right)  \ ,
\label{newa}
\end{eqnarray}
with $\lambda_3 >0$ and $\lambda_4 >0$. Of course, given the negative sign of the fourth order term in $H_1$, it is not anymore clear that the ground states of the full Hamiltonian $H$ are still regular graphs. If $\lambda_3$ and $\lambda_4$ are chosen small enough, however, it is plausible that $H_1$ becomes a small perturbation that just lifts the large original degeneracy and selects as ground states a particular subset of regular graphs. In this case, the positive third-order term clearly suppresses the formation of triangles in the regular graphs, while the fourth-order term ties up lots of links in the formation of squares. In the following I will show that this is indeed so. Before doing so, however, let me stress that, choosing $\lambda_3$ and $\lambda_4$ as small perturbations has the consequence that only very low-energy configurations are affected; the behaviour of the model at temperatures higher than the energy splitting caused by the perturbation remains the same and is dominated entirely by $H_0$. In the rest of this paper the values of the perturbative parameters will be chosen as $\lambda_3=0.01$ and $\lambda_4= 0.001$. 

Let me begin by deriving analytically the structure of the ground states for integer $d$, the cases on which I will henceforth concentrate, with the assumption that these are still regular graphs. I have already noted that the third-order term in $H_1$ prevents the formation of triangles. The fourth-order term, instead, favours the formation of squares. The ground state for integer $d$ will thus be a triangle-free $2d$-regular graph with the maximum number of squares. To derive what is the maximum number of squares in a uniform configuration I first observe that, by the degree sum formula $2e = \sum_{i\ge3} i \ v_i$, with $e$ the number of edges  and $v_i$ the number of vertices of degree $i$, one can derive that these graphs have exactly $dN$ edges. This means that one can uniquely assign to each vertex exactly $d$ edges. Out of $d$ edges one can form at most $d(d-1)/2$ squares. The ground state graphs will thus be triangle-free $2d$-regular graphs with $Nd(d-1)/2$ squares, each vertex having $d(d-1)/2$ squares uniquely assigned to it. Since a square is made of four vertices and four edges and there are a total of $N$ vertices and $dN$ edges, this means that each vertex is shared by exactly $2d(d-1)$ squares and each edge is shared by $2d-2$ squares. 

For simplicity, I will focus henceforth on the two cases $d=2$ and $d=3$, corresponding to connectivities four (as in Kazakov's model \cite{kazakov}) and six and describing space-times of geometric dimensions two and three, and I will compute the exact ground state energies in these cases. 

In the case $d=2$, the total number of squares is $S_2=N$. As derived above, each vertex is surrounded by four squares. The number of 4-cycles contributing to $H_1$ starting at each vertex is thus 32, 4 full squares traversed in two possible directions, for a total of 8 full squares, plus 4 back-tracking paths of two links per square, for a total of 16 paths, plus two "crosses" traversed in two possible directions, for a total of 4 "crosses", plus 4 double-back-tracking paths involving only one link, for a grand total of 32. This gives thus 
\begin{eqnarray}
S_2 &&= N \ , 
\nonumber \\
\left( H_0 \right)_2 &&= -{8\over 7} N \ .
\nonumber \\
\left( H_1 \right)_2 &&= -8\lambda_4 N \ .
\label{newb}
\end{eqnarray}
In the case $d=3$, instead there are $3N$ squares and each vertex is shared by 12 squares. An analogous counting argument for the number of 4-cycles gives
\begin{eqnarray}
S_3 &&= 3N \ , 
\nonumber \\
\left( H_0 \right)_3 &&= -{18\over 11} N \ .
\nonumber \\
\left( H_1 \right)_3 &&= -21\lambda_4 N \ .
\label{newb}
\end{eqnarray}

To conclude this section I would like to stress that the combinatorial properties just derived are identical with those of regular square and cubic lattices, although here these configurations are not posited but are, rather, dynamical since they appear as ground states of a Hamiltonian. The ground states of the Hamiltonian (\ref{newa}) are thus locally Euclidean graphs. In the next section I will analyze their combinatorial curvature properties. 

\section{Ricci curvature of the ground state graphs}
The first formulation of a discrete version of the Ricci curvature goes back to Regge \cite{regge}. What is, since then called Regge calculus, however is appropriate only for simplicial approximations of manifolds. Here we are dealing, instead with dynamical, emergent graphs. Two different but related versions of the Ricci curvature for generic graphs have been very recently proposed in the mathematical literature, the Ollivier combinatorial curvature \cite{olli1, olli2} and the Knill curvature \cite{knill}. As should be expected, both these quantities depend only on the neighbourhood of a particular vertex, the global topological properties of the ground state graph are irrelevant. As I now show, both notions of discrete Ricci curvature vanish for the ground states of the above model. 

As in the continuum the Ricci curvature is associated with a point and a direction on a manifold, both discrete versions of the same quantity are associated with a vertex $x$ and a link $e_x$ of a graph. As in the continuum limit, averaging over all links emanating from a vertex gives the discrete version of the Ricci scalar at that vertex. The difference between the Ollivier and the Knill discrete Ricci curvatures is that they try to capture and reproduce on discrete spaces two different features of the standard notion of continuum Ricci curvature. 

From a geodesic transport point of view, the Ricci curvature can be thought of as a measure of how much (infinitesimal) spheres (or balls) around a point contract (positive Ricci curvature) or expand (negative Ricci curvature) when they are transported along a geodesic with a given tangent vector at the point under consideration. The Ollivier curvature is a discrete version of the same measure. For two vertices $x$ and $y=x+e_x$ it compares the Wasserstein (or earth-mover) distance $W\left( \mu_x, \mu_y \right)$ between the two uniform probability measures $\mu_{x,y}$ on the spheres around $x$ and $y$ to the distance $d(x,y)$ on the graph and is defined as
\begin{equation}
\kappa_O (x,y)= 1- {W\left( \mu_x, \mu_y \right) \over d(x,y)} \ .
\label{ollia}
\end{equation}
The Wasserstein distance between two probability measures $\mu_1$ and $\mu_2$ on the graph is defined as
\begin{equation}
W\left( \mu_1, \mu_2 \right) = {\rm inf} \sum_{x,y} \xi(x,y)d(x,y) \ ,
\label{wasser}
\end{equation}
where the infimum has to be taken over all couplings (or transference plans) $\xi(x,y)$ i.e. over all plans on how to transport a unit mass distributed according to $\mu_1$ around $x$ to the same mass distributed according to $\mu_2$ around $y$, 
\begin{eqnarray}
\sum_y \xi (x,y) &&= \mu_1(x) \ ,
\nonumber \\
\sum_x \xi (x,y) &&= \mu_2(y) \ .
\label{transplan}
\end{eqnarray}
The Ollivier curvature is very intuitive but, in general not easy to compute. It is closely related to the number of triangles on the graph and thus to the Watts-Strogatz clustering coefficient \cite{graphrev}. Typically, the number of triangles touching a vertex constitutes a lower bound for the Ollivier curvature at that vertex \cite{liu}. For my purposes, however, the only important fact is the theorem that, for triangle-free graphs, the Ollivier curvature $\kappa_O (x,y)$ vanishes if and only if there is an exact matching between the unit-sphere neighbourhoods $N_x$ of $x$ and $N_y$ of $y$ \cite{olli2}, i.e. if there exists graph links that define a bijective relation that puts in one-to-one correspondence points in $N_x$ and in $N_y$. 

The ground state graphs of model (\ref{newa}) are indeed triangle-free. Moreover, I have derived in the previous section that in these ground state graphs each link is shared by exactly $2d-2=k-2$ squares, where $k$ is the connectivity. Let me consider thus the link between two neighbouring vertices $x$ and $y$. A square based on this link comes with another link joining a neighbour of $x$ with a neighbour of $y$. Since there are exactly $k-2$ such squares, this means that there are unique links joining $k-2$ neighbours of $x$ to $k-2$ neighbours of $y$. Apart from the original link joining $x$ and $y$ this leaves out exactly one link to an additional neighbour of $x$ and another one to an additional neighbour $y$. But the former defines a link between a neighbour of $x$ and $x$, which is itself a neighbour of $y$ and the latter defines a link between $y$, which is the last remaining neighbour of $x$ and the last remaining neighbour of $y$. Each link of the graph being shared by exactly $k-2$ squares is thus the necessary and sufficient condition for the existence of an exact matching between the respective unit sphere neighbourhoods of all pairs of neighbouring vertices. The ground state graphs derived in the previous section are thus Ricci-flat graphs according to the Ollivier measure.

The second approach to a discrete version of the Ricci curvature for graphs is due to Knill \cite{knill} and bases on the fact that continuum Ricci curvature of a manifold is the sectional average of the curvatures over all two-dimensional sections in the tangent space to the manifold at a given point and that contain a particular tangent vector. The discrete curvature of a graph at vertex $x$ is given by 
\begin{equation}
K(x) = \sum_{k=1}^{\infty} \ (-1)^k {V_{k-1} (x) \over k+1} \ ,
\label{gauss}
\end{equation}
where $V_k (x)$ is the number of $K_{k+1}$ complete subgraphs in the unit sphere $S(x)$ (the neighbourhood) at vertex $x$. This means $V_0$ is the number of vertices in the neighbourhood of $x$, $V_1$ the number of edges in the same neighbourhood, $V_2$ the number of triangles and so on. Given a vertex $x$ and an edge $e$, the discrete version of two-dimensional sections at $x$ containing $e$ is taken as the ensemble of wheel graphs $W_n$, $n\ge 4$ that contain $e$ as an edge connected to the center. Wheel graphs $W_n$ are made of a cycle $C_n$ connected to the vertex $x$ at the center. The restriction $n\ge 4$ excludes the tetrahedron $W_3$ since this is not a two-dimensional geometric graph (see below). The first two-dimensional structure that "feels" curvature is thus $W_4$ which is made of 4 adjacent triangles. Ricci curvature $\kappa_K (x, e)$ is then taken as the average over the curvatures of all wheel graphs $W_n$, $n\ge 4$ that have $x$ as center and $e$ as an edge connecting the center to the cycle. It is then clear that graphs that contain no triangles have vanishing curvature according to this definition, since they also contain no wheel subgraphs. The ground state graphs of model (\ref{newa}) are thus Ricci-flat also according to the Knill measure. 

While the ground states are Ricci flat, it is easy to convince oneself that they admit local excitations corresponding to curvature quanta. First of all, let me note that, from (\ref{gauss}), the Ricci curvature of wheel graphs is given by
\begin{equation}
K\left( W_n \right) = 1-{n\over 6} \ .
\label{curwheel}
\end{equation}
The two fundamental quanta of curvature are thus $\pm 1/6$ and correspond to the wheel graphs $W_5$ and $W_7$, $W_6$ being flat. As I have shown above, for both $d=2$ and $d=3$, every vertex is shared by at least 4 squares. Let me thus consider such a configuration of 4 squares touching at a central vertex. The elementary local quanta of positive and negative curvature are then obtained by adding 4 additional links that involve only the first- and second-nearest neighbours of the central vertex, as shown in Figs. 1 and 2. 

\begin{figure}
\includegraphics[width=8cm]{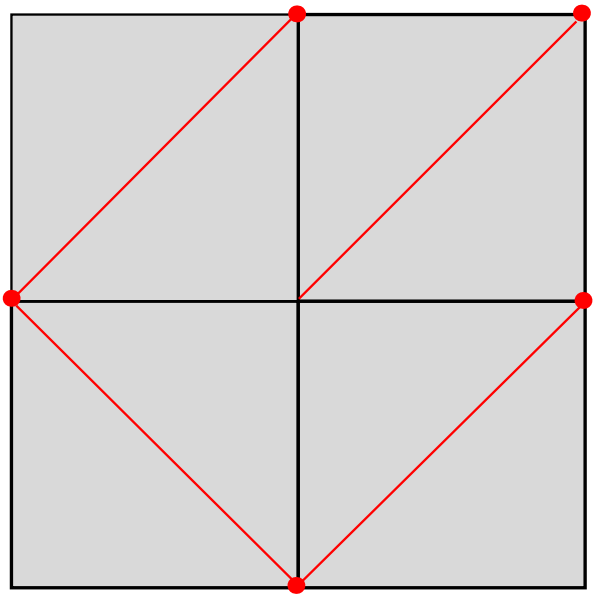}
\caption{\label{fig:Fig. 1} The positive curvature quantum corresponding to the wheel graph $W_5$. In red (oblique lines) the additional links making up the excitation.}
\end{figure}

\begin{figure}
\includegraphics[width=8cm]{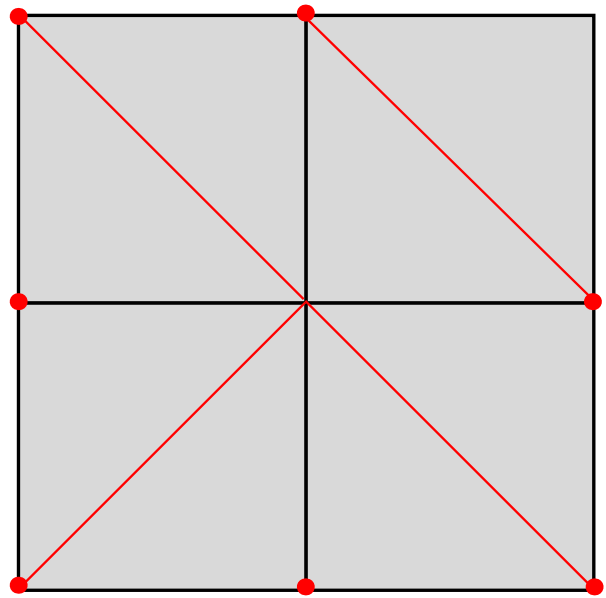}
\caption{\label{fig:Fig. 2} The negative curvature quantum corresponding to the wheel graph $W_7$. In red (oblique lines) the additional links making up the excitation.}
\end{figure}

The energies of these curvature quanta can be computed by a tedious, but straightforward counting of the new interactions introduced by the additional links. For example, for $d=2$ they are given by 
\begin{eqnarray}
E_{d=2} \left( W_5 \right) &&= {6\over 7} +16  \lambda_3 - {55\over 2} \lambda_4 \ ,
\nonumber \\
E_{d=2} \left( W_7 \right) &&= 1 +16  \lambda_3 - {55\over 2} \lambda_4 \ .
\label{enerquanta}
\end{eqnarray}
Interestingly, the negative curvature quantum costs more energy than its positive curvature counterpart. This is because, for $W_7$, three of the additional links meet at the same point (see Fig.2), whereas this is never the case for $W_5$ (see Fig.1). 

\section{Random large worlds with emergent dimensions}
Suppose we draw the same four-regular ($2d$) graph on two pieces of paper. The local geometric structure determines its energy according to the Hamiltonian (\ref{newa}): as I have shown above, this depends only the local number of edges, triangles and squares. Suppose we now leave the first piece of paper flat while we crumple the second one up like when using a newspaper page to light a fire. The local geometry and thus the energy of the two graphs remain clearly the same. The topology, however changes dramatically. The first remains flat while the second becomes space-filling and acquires thus one additional (Hausdorff) dimension: there is a large topological degeneracy. 

The topological structure depends thus not only on the Hamiltonian but also on how the fluctuations within the degenerate ground state manifold are sampled. As in \cite{tru1} I will consider quenching \cite{quench} the graphs from very high energy random configurations by a Glauber dynamics. It is well known that quenching can lead to complex behaviour when various phases can coexist at low temperatures. In many cases the equilibrium configuration is not always reached (e.g. only in 70\% of the cases for the two-dimensional Ising model) and in some other cases (three-dimensional Ising model) it is never reached \cite{quench}. Here I will study the properties of the lowest-energy graphs obtained by quenching random configurations for the coupling constant values $d=2$ and $d=3$. These can be considered as the configurations resulting from rapid cooling (or rapid expansion via the renormalization group) a system governed by the Hamiltonian (\ref{newa}). 

To do so I will adapt the Glauber dynamics introduced in \cite{tru1} to take into account the perturbative modifications to the Hamiltonian. So I will start from random initial configurations, with typically very high-energy, and sequentially update the vertex and link bits according to the rule
\begin{eqnarray}
&&s_i (t+1) = {\rm sign}_+ \left( h_i(t) \right) \ , \nonumber \\
&&h_i = \sum_{j\ne i} w_{ij} s_j  \ ,
\label{twof} \\
&&w_{ij}(t+1) = \Theta _{\pm} \left( h_{ij} \right) \, \nonumber \\
&&h_{ij} =  {1\over 2} s_i  s_j  - {J\over 2} \sum_{k\ne j \atop k\ne i} w_{kj}(t) - {J\over 2} \sum_{k\ne i \atop k\ne j} w_{ik}(t) 
\nonumber \\
&& \ \ \ \ \ -\lambda_3 \left( w^2 \right) _{ij} + \lambda_4 \left( w^3 \right) _{ij} \  ,
\label{twog}
\end{eqnarray}
where $\Theta$ denotes the Heaviside function. The subscripts "+" and "$\pm$" on the "sign" and $\Theta$ functions indicate what is the rule to follow in the (rare) cases in which the argument happens to be zero. Complete symmetry would require a totally random choice in these cases. This, however can lead to a split, two-component network in which space-time vertices have the opposite sign in the two components. Since I am interested in the emergent properties of just one of these components I will maximize its size by making an asymmetric choice for the "sign" function determining the orientation of the space-time vertices. This is embodied in the subscript "+", which indicates that, in case of vanishing argument the + sign has to be chosen. This favours positive vertex spin values, so that one-component graphs are preferred. The subscript "$\pm$" on the Heaviside function, instead indicates that a purely random choice will be made when the argument vanishes. 

The important property of the above Glauber dynamics is that the energy function ({\ref{newa}) cannot increase along a sequential evolution (\ref{twof}) and (\ref{twog}): $E(t+1) \le E(t)$. To show this, let me begin by considering the update of vertex spin $s_i$. The corresponding contribution $E_i$ to the energy changes according to
\begin{eqnarray}
E_i(t+1) &&= - s_i(t+1) \sum_{j\ne i} w_{ij} (t) s_j(t) 
\nonumber \\
&&= - {\rm sign}_+ \left( h_i(t) \right) h_i(t) =- |h_i(t)| 
\nonumber \\
&&\le -s_i(t) h_i(t) = E_i(t) \ .
\label{twoh}
\end{eqnarray}
With the exact same procedure for the update of a link spin $h_{ij}$ I obtain
\begin{equation}
E_{ij}(t+1) = - w_{ij} (t+1) h_{ij} (t) = -\Theta_{\pm} \left( h_{ij} (t) \right) h_{ij}(t) \ .
\label{twoi}
\end{equation}
Now there are two possibilities 
\begin{itemize} 
\item $h_{ij} (t) <0$ $\rightarrow$ $E_{ij}(t+1) =0 \le -w_{ij}(t)  h_{ij}(t) = E_{ij} (t)$ 
\item $h_{ij} (t) > 0$ $\rightarrow$ $E_{ij}(t+1) =- h_{ij}(t) = -|h_{ij} (t) | \le -w_{ij}(t)  h_{ij}(t) = E_{ij} (t)$ 
\end{itemize}
and in both cases we obtain $E_{ij} (t+1) \le E_{ij}(t)$, which proves the claim that the energy cannot increase during sequential evolution according to the above Glauber dynamics. Note that, for both the "sign" and the Heaviside functions, the exact procedure on how to treat the undefined cases in which the argument vanishes has no effect on this result. As a consequence, every minimum of the energy (\ref{newa}) is a fixed point of the sequential network evolution (\ref{twof}) and (\ref{twog}). 

Once the adjacency matrix $w$ of the ground state has been obtained, two distance measures on this graph can be studied. 
The spectral dimension $d_s$ measures the connectivity of the graph \cite{spectral}, the dimension that a particle moving on the graph would feel.  It is defined via the scaling of the return probability $p_r(t)$ to the initial point after $t$ steps of a random walk on the graph \cite{rwalk},
\begin{equation}
p_r (t) \propto t^{-d_s / 2 } \ .
\label{twoj}
\end{equation}
For infinite graphs this scaling relation is valid in the limit $t\to \infty$. For finite $k$-regular graphs, the return time to the initial point is $t=N$ \cite{rwalk} and thus $p_r(t) \to 1$ in the limit $t \to \infty$. The correct scaling is typically found in the intermediate region $1\ll t \le O(N^{4/k})$ where finite size effects are suppressed \cite{rwalk}. To compute the spectral dimension one introduces the degree distribution $d_i$ denoting the number of edges at each vertex $i$. The matrix
\begin{equation}
M_{ij}  = {1\over d_i} A_{ij} \ ,
\label{specdima}
\end{equation} 
represents the transition probabilities from vertex $i$ to vertex $j$ in a random walk on the ground state graph. The average over all possible initial points of the return probability of the random walk after $t$ steps is then given by
\begin{equation}
p_r (t) = {1\over N} \ {\rm Tr} M^t \ .
\label{specdimb}
\end{equation}
The best fit of this to a power $t^{-d_s / 2 }$ over the appropriate $t$-interval defines then the spectral dimension $d_s$. In practice, I have computed $d_s$ by using the five points $t_{\rm max} \dots t_{\rm max}-4$ with $t_{\rm max} = {\rm ceiling} (N^{4/k})$ for the lowest-energy quenched graph with the largest $N$. This gives
\begin{eqnarray}
d_s &&= 2 \pm 0.02 \ \qquad  \ \ d=2 \ ,
\nonumber \\
d_s &&=3.02 \pm 0.02 \ \qquad d=3 \ ,
\label{specdim}
\end{eqnarray}
which confirms that, indeed, the coupling constant of the model determines the spectral dimension of the ground state graphs. 

The intrinsic Hausdorff dimension $d_H$, instead, measures how the graph volume scales with distances ${\cal D}$ on the graph \cite{graphrev},
\begin{equation}
{\cal D} \propto N^{1/d_H} \ .
\label{twok}
\end{equation}
Of course, one must specify the exact definition of the graph distance, which I will address below. But the asymptotic results do not depend on the details of this definition, being a topological property of the graph. 

Denoting $D(q,p)$ as the distance between vertices $p$ and $q$ on the graph $G$, i.e. the number of edges on the shortest path connecting them, let me define the sphere $S$ and ball $B$ of radius $r$ around a vertex $p$ as
\begin{eqnarray}
S_p(r) &&= \{ q \in G | D(q,p)=r \} \ ,
\nonumber \\
B_p(r) &&= \{ q \in G | D(q,p)\le r \} \ .
\label{newsphereball}
\end{eqnarray}
Clearly, we have
\begin{equation}
B_p(R) = \sum_{r=0}^R S_r \ .
\label{sumspheres}
\end{equation}
Let me now consider a vertex on $S_p(R)$, i.e. a vertex at distance $R$ from $p$ and let me ask the question how many unconstrained ways there are to proceed to a next vertex at distance $R+1$ from $p$. There are two extreme cases for regular graphs. The first is the case of a complete graph, for which each vertex is at distance one from $p$ and, thus $S_p(r) = (N-1) \delta_{r,1}$ and $B_p(R) =N$ independent of $R$ for $R>1$, which amounts to a vanishing Hausdorff dimension. The second case is that of a random regular graph for which edges are drawn randomly from a uniform distribution. In this case, for each vertex on $S_p(R)$ there are exactly $(k-1)$ independent ways to proceed to a new vertex, which means that $B_p(R) \propto (k-1)^R$ for large $R$. This, in turn implies that distances on the graph scale logarithmically with the volume ("infinite Hausdorff dimension"), which is what goes under the name of "small world" phenomenon \cite{graphrev}. 

The ground state graphs of (\ref{newa}) constitute an intermediate case, in between these two extremes. Indeed, consider as before a vertex on $S_p(R)$ and the link joining it to a previous vertex on $S_p(R-1)$. As I have shown before, this link is shared by $(k-2)$ squares and this means that at least $(k-2)$ of the new links emanating from the vertex on $S_p(R)$ have to be tied up in forming these squares. In the extreme case that all these $(k-2)$ links are constrained to complete squares involving vertices already in $B_p(R)$ there remains only one, free, unconstrained link to to further continue the path from the vertex on $S_p(R)$. In other words, in this case $S_p(R)$ is a constant and $B_p(R) \propto R$, which amounts to a Hausdorff dimension one. In general, however, there will be more free, independent links to continue the path since not all of the $(k-2)$ links emanating from a point on $S_p(R)$ will be constrained to close squares by turning back into $B_p(R)$. The constraint of forming $(k-2)$ squares per link, however, clearly prevents all the links emanating from boundary points to be free to continue the path to $S_p(R+1)$. In general, for $S_p(r)$ scaling like $r^{d_H-1}$ one obtains $B_p(R)\propto R^{d_H}$ by Faulhaber's formula. This shows that, indeed, the constraint of maximizing the number of squares (with no triangles) rules out a logarithmic scaling of graph distances as a function of the volume in favour of a power law. The ground states of (\ref{newa}) are thus "large worlds" with finite Hausdorff dimension. 

As mentioned before, however, it is possible that the quenched graphs are not exact ground sates of (\ref{newa}) but only configurations "near" such a ground state. In order to find out what exactly is the Hausdorff dimension of the quenched graphs, I will fit (\ref{twok}) to the numerical results obtained from the Glauber dynamics (\ref{twof}) and (\ref{twog}). Following \cite{jonsson} I will take ${\cal D} =\langle D \rangle$ as the average distance among vertices of the graph. This can be also computed from the adjacency matrix $w$ of the ground state graph. Indeed, the minimal graph distance between two vertices $i$ and $j$ is given by the smallest integer power ${\ell}_{ij}$ such that $\left(w^{{\ell}_{ij}}\right)_{ij} \ne 0$. The average distance on the graph is thus given by
\begin{equation}
\langle D \rangle = {1\over N(N-1)} \sum_{ij} {\ell}_{ij} \ .
\label{hau}
\end{equation}
Let me stress that this is an intrinsic property of the graph that does not need an embedding in an extrinsic Euclidean space to be defined and measured. This dimension is simply extracted as the best fit of (\ref{hau}) to a power of $N$ as in (\ref{twok}). 

In order to measure the Hausdorff dimension of the quenched graphs I have taken the mean $\ll {\cal D} \gg $ of the average graph distances, as defined above, for the 3 graphs of lowest energy obtained by the quenching process and have fit the results to a functional form
\begin{equation}
\ll {\cal D} \gg = c\  N^{1/d_H} \ .
\label{haus}
\end{equation}
Here, $\ll \dots \gg$ denotes averages over graphs of low energies as opposed to averages over properties of a single graph. The measured values of ${\rm log} \left( \ll {\cal D}\gg \right)$  are shown in Fig. 3 and Fig. 5 as a function of ${\rm log}(N)$ for $d=3$ and $d=2$, respectively. The values of $N$ are limited by below by the finite-size effects appearing in too small graphs and by above by the available computational resources. 

\begin{figure}
\includegraphics[width=8cm]{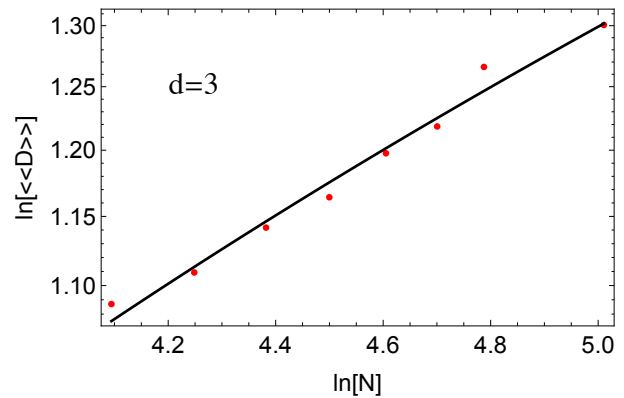}
\caption{\label{fig:Fig. 3} The linear regression of eq. (\ref{haus}) for $d=3$ in a ln-ln plot. }
\end{figure}

In the case $d=3$ the linear regression of these logarithmic data (shown in Fig. 3) implies best fit parameters
\begin{eqnarray}
c &&= 1.05 \pm 0.07 \ ,
\nonumber \\
d_H &&= 4 \pm 0.2Ê\ .
\label{fitthree}
\end{eqnarray}
The small error on $d_H$ of only about 5 \% strongly suggests that the Hausdorff dimension of the quenched graphs is $d_H=4$. Note that the Akaike information criterion value of the power-law fit is ${\rm AIC} ({\rm power\ law} )=-26.4 $, while, for comparison, the Akaike information criterion of a logarithmic fit $\ll {\cal D} \gg =c \  {\rm log}(N) $ is ${\rm AIC} ({\rm logarithm} )= -23.2$. This implies that the logarithmic fit is 5 times less likely to be the best model (in the sense of minimizing the information loss) for the measured data, confirming that the quenched graphs are "large worlds". Indeed, as shown in Fig. 4, the quenched graphs are actually locally Euclidean ground states.

\begin{figure}
\includegraphics[width=8cm]{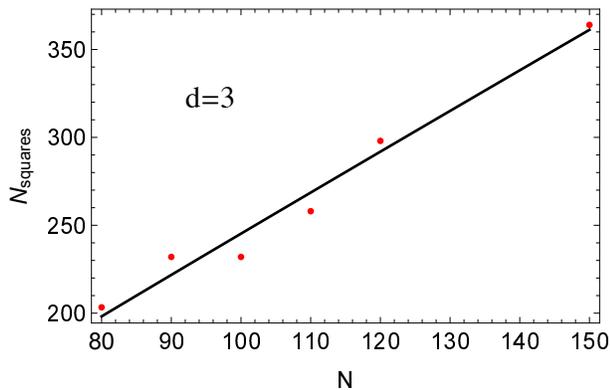}
\caption{\label{fig:Fig. 4} The number of squares in the quenched graphs for $d=3$ as a function of $N$.}
\end{figure}
The best fit of the number of squares $N_{\rm squares}$ in the quenched graphs as a function of $N$ gives 
\begin{eqnarray}
N_{\rm squares} &&= p \ N^q \ ,
\nonumber \\
p &&= 3 \pm 1.1\ ,
\nonumber \\
q &&= 0.96 \pm Ê0.08\ ,
\label{fitthreesquares}
\end{eqnarray}
which are essentially the values of cubic graphs, albeit with a large error in the coefficient $p$ with the available precision. 

\begin{figure}
\includegraphics[width=8cm]{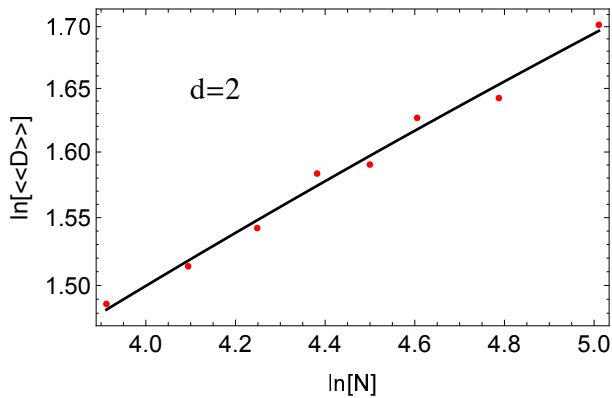}
\caption{\label{fig:Fig. 5} The linear regression of eq. (\ref{haus}) for $d=2$ in a ln-ln plot. }
\end{figure}

The situation is quite different for $d=2$. In this case, the best fit parameters implied by the linear regression of the logarithmic data in Fig. 5 are
\begin{eqnarray}
c &&= 2.05 \pm 0.08 \ ,
\nonumber \\
d_H &&= 5.1 \pm 0.2Ê\ .
\label{fittwo}
\end{eqnarray}
The error on $d_H$ is again only of 4\%. This result strongly suggests that, for $d=2$ the Hausdorff dimension of quenched graphs is actually $d_H=5$. Also in this case one can exclude a "small world" behaviour since the Akaike information criterion of the power law is ${\rm AIC} ({\rm power\ law} )= -24.3$, compared with a value of (-17.6) for ${\rm AIC} ({\rm logarithm} )$, which implies that a logarithmic fit is 28 times less likely to be the best fit of the data (in the sense of minimizing the information loss) than a power law. Also for $d=2$, quenched graphs are "large worlds". As extensively discussed in section VII, however, they are not locally Euclidean since the squares are much too sparse. 

In order to explain how the Hausdorff dimensions $d_H=4$ and $d_H=5$ come about, I will introduce geometric graphs and the notion of geometric dimension, which is a measure related, but not identical to the Hausdorff dimension. Geometric graphs are discrete versions of topological manifolds, manifolds that are locally Euclidean since each point has a neighbourhood homeomorphic to an open set in $\mathbb{R}^n$. For even connectivities ($d$ integer), the triangle-free, $2d$-regular graphs with maximum number of squares considered here are clearly geometric since, as I have shown above, each edge supports ($2d-2$) square, i.e. has a neighbourhood homeomorphic to $\mathbb{Z}^d$, which is a natural discretization of $\mathbb{R}^d$. 

To define geometric graphs in the generic case \cite{knill} one must require the possibility to define unit spheres $S(x)$ on the graph with the same topological properties as in topological manifolds. Thus a graph is called a geometric graph of geometric dimension $d_g$, if the dimension is constant for every vertex $x$ and if each unit sphere S(x) is a $(d_g -1)$-dimensional geometric graph with Euler characteristic 
\begin{equation}
\chi (S(x)) = 1 - (-1)^{d_g} \ ,
\label{euler}
\end{equation}
where the Euler characteristic $\chi (S(x))=\sum_{x\in S(x)} K(x)$ is the sum of the curvatures defined in (\ref{gauss}); a one-dimensional graph is geometric if every unit sphere has Euler characteristic 2 \cite{knill}. For $d_g=1$ the minimal geometric graph is the square, for $d_g=2$ it is the octahedron (the icosahedron is also a $d_g=2$ geometric graph) while for $d_g=3$ it is the 16-cell (also called hexadecachoron or hexadecahedroid). In general, for any dimension $d_g$ the minimal geometric graph is the cross polytope of dimension $D_g=d_g+1$, whose vertices consist of all permutations of $(\pm 1, 0, \dots ,0)$ in $D_g$ Euclidean dimensions. Note that generic triangularizations of manifolds do not yield geometric graphs since, without further assumptions, the unit spheres can be arbitrarily complicated because pyramid constructions allow any given graph to appear as a unit sphere of an other graph \cite{knill}. In particular triangles and tetrahedra are not geometric graphs. 

For a single elementary cross polytope it is of course a matter of convention if one calls geometric dimension $d_g$ or $D_g=d_g+1$. If the cross polytope, however is embedded as a subgraph in a larger, regular graph, its geometric dimension must be considered $D_g$, exactly as 
embedding a continuum unit sphere in a larger topological manifold requires at least one more dimension. 

Let me consider some examples. Consider first an octahedron, i.e. a $D_g=3$ cross polytope. Each vertex of the octahedron has 4 links emanating from it: if an octahedron has to be embedded in a larger graph one needs thus at least some vertices with connectivity 5. Graphs with vertices of connectivity 6 also generically contain octahedrons, for examples octahedron balls in which the center is linked to all vertices on the boundary. They do not, however contain 16-cells: for this, vertices of connectivity 7 are needed since each vertex of the 16-cell has 6 links emanating from it. So, regular graphs of connectivity $k=3$ or $k=4$ can be assigned geometric dimension $D_g=2$, regular graphs of connectivity $k=5$ or $k=6$ have geometric dimension $D_g=3$ and regular graphs of connectivity $k=7$ or $k=8$ geometric dimension $D_g=4$. In general, regular graphs of connectivity $k$ can be assigned a geometric dimension $D_g = {\rm ceiling} (k/2)$. This encompasses also the present case of $2d$-regular graphs locally homeomorphic to $\mathbb{Z}^d$.

Let me now turn to the quenching process of high-energy configurations via the Glauber dynamics (\ref{twof}) and (\ref{twog}) and concentrate on the link update (\ref{twog}) assuming a fully formed space, $s_i=1$, $\forall i$. If one assumes a very high-energy configuration with a macroscopic portion (I will take 50\% in the numerical simulations below) of links present ($w_{ij}= 1$), the first step of the Glauber dynamics will set most links to zero since most of the $h_{ij}$ are negative due to the high values of the second and third negative terms in (\ref{twog}) (the remaining fourth and fifth terms are irrelevant since they constitute by construction only perturbations). But then, the second step will again generate many links for exactly the opposite reason, the first term due to the fully formed space will dominate and cause most of the $h_{ij}$ to take a positive value. The quenching process will thus approach the final regular graph by alternating between configurations with many and few links. 

In Fig. 6 I display the average connectivity and the number of links that remain unchanged in the next step of the quenching process as a function of quenching time $t$ for $d=3$, starting from a high-energy configuration with 50\% of all possible links randomly set to 1. 

\begin{figure}
\includegraphics[width=8cm]{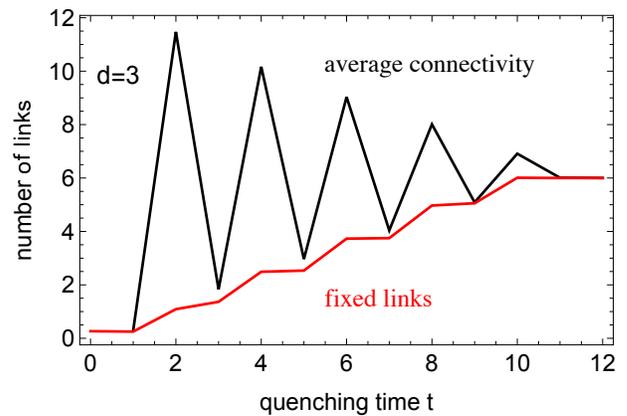}
\caption{\label{fig:Fig. 6} The average connectivity (upper graph) and the number of links that remain unchanged to the next step (lower graph) in the quenching process for $d=3$.}
\end{figure}
In this figure one can clearly recognize the approach to the final connectivity 6 by alternating between configurations with large and small numbers of links. Also note that, at the beginning of the quenching process, the links are completely random variables while, slowly, as the quenching proceeds, some of them begin to remain fixed from one step to the other. Since, for $d=3$, the quenched graphs are locally Euclidean, I will focus on their geometric dimension as defined above. The first quenching step in which a graph of geometric dimension 3 is fully formed is when the average number of links that remain unchanged reaches the level 5: as is clear from Fig. 6 this happens at $t=8$, when the average number of fixed links reaches the level 4.97. The important point is that, exactly at this quenching time, the average connectivity is 8.01, with essentially all vertices having connectivity 8. This means that, at this time we have a fully formed graph of geometric dimension 3 embedded in a larger, locally Euclidean graph of geometric dimension 4: at each vertex the links of the geometric dimension 3, quenched graph can "choose" among the 8 links of a geometric dimension 4, spontaneously emerged embedding graph. Since only the last link per vertex will be adjusted in the last few steps of the quenching process, the "choice of directions" made at step 8 remains "frozen" and this causes the final graph to be a discrete, four-volume-filling, three-dimensional space, which explains its Hausdorff dimension 4. The quenching process has generated a spontaneous embedding graph of geometric dimension 4: the quenched ground state graph is embedded in this by a local, random choice of 6 of the 8 links of the embedding graph at each vertex. At each vertex there are thus two "missing" links, i.e. one entire, locally Euclidean dimension: this is the spontaneous dimension emerged in the quenching process. 

The situation is more complex for $d=2$, for which case the behaviour of the average connectivities and of the fixed number of links in the quenching process are shown in Fig. 7. 

\begin{figure}
\includegraphics[width=8cm]{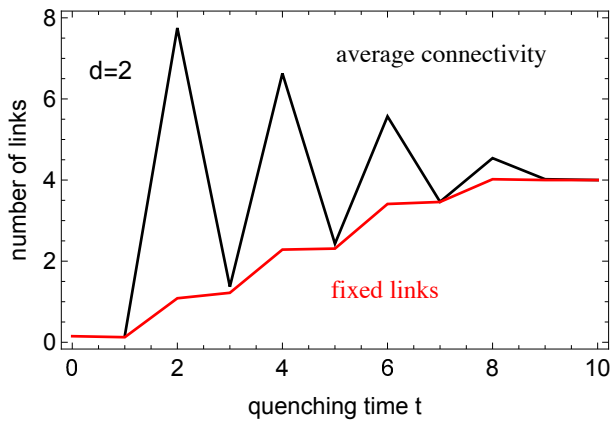}
\caption{\label{fig:Fig. 7} The average connectivity (upper graph) and the number of links that remain unchanged to the next step (lower graph) in the quenching process for $d=2$.}
\end{figure}
As the quenched graphs are not locally Euclidean, as discussed in section VII, one cannot use anymore the notion of geometric dimension to determine when, in the quenching process, the final configuration of spectral dimension $d_s=2$ has formed. 

\section{Emergent Time and Causality}
In the previous section I have presented ample evidence that quenched networks of sufficiently low energy for $d=3$ are locally Euclidean, "four-space-filling" (from now on denoted by $4D$), three-dimensional (denoted as $3d$) graphs. Here I would like to derive an important  topological property of such configurations.  

The quenched ground states are essentially cubic-like networks defined on a regular graph with connectivity $k=8$. These, on the other side, can be viewed as cubic-like networks with two additional "missing links" at every vertex. These links define a fourth, emergent dimension. This is of course very suggestive of a possible mechanism for the appearance of a time direction. In the following I will explore if such an identification satisfies the necessary requirements to be consistent. The first thing to notice is that the additional links at each vertex define a discrete version of a nowhere vanishing direction field on a continuous manifold. A nowhere vanishing direction field is the necessary and sufficient condition for a Lorentz metric to exist on a manifold, i.e. for the manifold to be Lorentzian \cite{geroch}. Indeed it is very easy to construct this Lorentzian metric from the standard Euclidean metric on the manifold \cite{hawking}. Most important, this construction carries through to discrete graphs: it suffices to take the Euclidean distance along the shortest path between two points on the graph and to substract twice the Euclidean distance along the additional links on this path. This reduces to the standard procedure described in \cite{hawking} in the continuum limit. 

A Lorentzian distance is the first requirement for a space-time. The second is the existence of a global definition of time, i.e. the possibility of defining a global orientation on the direction field, or in other words, the arrow of time. To see that this poses no problem it is best to think of the additional two links per vertex as defining a new regular graph with connectivity two. It is well known that two-regular graphs decompose in many disjoint cycles (closed loops) \cite{graphrev}. For each of these loops one can then independently choose one of the two possible orientations in which the cycle should be traversed and this choice defines automatically a global orientation of time without conflicts, since the cycles are disjoint. 

The final and most difficult requirement for a physical space-time is causality. There are many types of causality that have been considered \cite{minguzzi} for continuous manifolds. Here I will concentrate on one of the simplest notions of causality, i.e. the absence of closed causal curves. To this end let me consider in the full 4D graph a closed path (cycle) of length $L=O\left( N^{1/4}\right)$ made of $s$ steps along the 3d ground state graph and the rest along the additional emergent "time" direction. In particular I will ask the question of how many such cycles there are in the limit of very large $N$. To answer this question I will use two fundamental results about the number of cycles in regular graphs. The first \cite{loopreg1} is that, for connectivity $k\ge 3$ , the number $C$ of closed paths (cycles) of length $L=o(N)$ in a k-regular graph is given by 
\begin{equation}
C = {(k-1)^L \over 2L} \ ,
\label{ncrg}
\end{equation}
The second, very recently derived \cite{loopreg2}, is that this formula holds for all connectivities, including $k=2$. This shows that every vertex along the path contributes an overall factor $(k-1)/{\rm log}_L(2L) $ where $(k-1)$ are the number of possibilities of how to continue the path once arrived at that vertex. The total number of cycles with $s$ steps along the 3d ground state graph is thus given by
\begin{equation}
C = {(k-1)^s \over 2L} \ ,
\label{ncst}
\end{equation}
with $k=2d=6$. 

In the limit $N\to \infty$ there are thus only two possible regimes:
\begin{itemize}
\item{} $s = o\left( {1\over 4} {\rm log}_{k-1}(N) \right)$,
\item{} otherwise, 
\end{itemize}
the extreme cases for the two regimes being given by the integers 
\begin{equation}
s_- ={\rm floor} \left( {1\over 4} {\rm log}_{k-1}(N) \right) \ ,
\label{newflo}
\end{equation}
and
\begin{equation}
s_+= {\rm ceiling} \left( {1\over 4} {\rm log}_{k-1}(N) \right) \ ,
\label{newceil}
\end{equation}
respectively. In the limit $N\to \infty$ no cycles with $s \le s_-$ survive. I will thus identify 4D graph paths with $s \le s_-$ as causal paths on the full 4D graph. To end this section I would like to stress that, in this model the existence of a dimension interpretable as time would be a consequence of the quenching process, i.e. of a rapid cooling associated with a sudden expansion to large scales, and that the ensuing notion of causality would be emergent, appearing only at large distances. This picture would entail, in particular, that time arises as an ``antiferromagnetic configuration for directions" of a fundamentally discrete Euclidean 4D space. I want to stress, however that, while I have shown how it is possible to define a globally well defined "Lorentzian graph distance" with a causal structure, there is no compelling evidence in the present framework why this should be the correct choice. 

\section{Graph holography and minimal areas/loops as fundamental degrees of freedom}
Let me now proceed to the coupling $d=2$. In this case, the quenched graphs of lowest energy are four-regular graphs of spectral dimension $d_s=2$ but Hausdorff dimension $d_H=5$. As before, this means, loosely speaking, that the quenched graphs are two-dimensional structures somehow embedded in a larger, five-dimensional discrete space. But in this case the identification of this space is harder than for $d=3$. 

To see why, let me first note that the mismatch of dimensions has important consequences for the counting of degrees of freedom. Indeed, since the quenched graphs are two-dimensional structures, their fundamental degrees of freedom must have a finite density proportional to the inverse area element on the graph and their total number $N_{\rm dof}$ scales thus as $N_{\rm dof} \propto {\cal D}^2 \propto N^{2/5}=N^{0.4}$, given that distances on the two-dimensional quenched graphs scale with Hausdorff dimension 5. This means that the original vertices of the graph constitute a redundant representation of the true, physical degrees of freedom: there are only $N^{2/5}$ of these, a massive reduction in the large $N$ limit. This is a graph version of the holographic principle \cite{holo}, perhaps the most established and generally accepted feature of quantum gravity. 

But can we identify a possible candidate for these physical degrees of freedom? The answer is yes and to derive it I am going to count the number of squares in the quenched graphs as a function of the number $N$ of vertices. For $d=3$ this gives $N_{\rm squares} = 3 N$ as derived in (\ref{fitthreesquares}) and shown in Fig. 4 and this is the expected number for locally Euclidean graphs. This is not so for $d=2$, as shown in Fig. 8. 

\begin{figure}
\includegraphics[width=8cm]{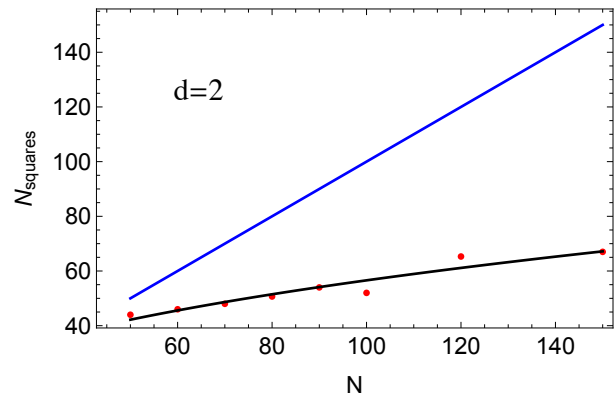}
\caption{\label{fig:Fig. 8} Graph holography: the number of squares for $d=2$ quenched graphs scaling as $N^{2/5}$ versus the linear behaviour characteristic of locally Euclidean graphs. }
\end{figure}

The best fit of the behaviour shown in Fig. 8 gives
\begin{eqnarray}
N_{\rm squares} &&= p \ N^q \ ,
\nonumber \\
p &&= 8 \pm 1.9\ ,
\nonumber \\
q &&= 0.42\pm Ê0.05\ .
\label{fittwosquares}
\end{eqnarray}

While the present data do not permit to determine with high accuracy the coefficient, the exponent is obtained with a much better error of 12\%. This suggests that the number of squares actually scales with an exponent $0.4 = 2/5$. There are three immediate consequences of this result. First, the quenched graphs are clearly not locally Euclidean, since squares are much too sparse: this makes the interpretation of Fig. 7 harder for $d=2$, since one cannot clearly identify when a two-dimensional structure has formed in the quenching process. The identification of the spontaneous embedding space and its properties for $d=2$ is thus still an open question. Secondly, given this sparsity of squares there is typically no more a perfect matching between unit spheres of neighbouring vertices. The Ollivier Ricci curvature of the quenched graphs is thus typically negative \cite{liu} (the Knill Ricci curvature still vanishes because of the absence of triangles but it is a question if it is appropriate for these non locally Euclidean graphs). Finally, squares on the quenched graphs are obvious candidates for the fundamental degrees of freedom of the model. Since the quenched graphs are, by construction, triangle-less, squares are the fundamental quanta of area and constitute the smallest possible loops. In the present model, holography is thus realized by fundamental degrees of freedom that are quantized loops/areas on a space-filling 2d holographic screen embedded in a 5D space of negative curvature. By taking averages over suitably large subgraphs one can expect to obtain a constant negative curvature, in which case the natural embedding would become the hyperbolic space $\mathbb{H}^5$, which is topologically $\mathbb{R}$ x $\mathbb{R}^4$ with a warped metric.

\section{Conclusion and Future directions} 
In this paper I have proposed a model that might give a first hint towards a possible explanation for the spontaneous emergence of a time direction, namely the quenching of graphs governed by link antiferromagnetism as in (\ref{newa}). The mechanism is easiest to present for $d=3$, in which case quenching does not destroy the locally Euclidean nature of the ground states. For $d=2$ the coupling is stronger and curvature effects become more important. In this case the quenched graphs are space-filling 2d holographic screens in a 5D space of negative curvature and the fundamental degrees of freedom are quantized loops/areas on this holographic screen. It is interesting to note that five dimensions play a crucial role in the AdS/CFT scenario \cite{maldacena} and that quantized areas are one of the most important aspects of loop quantum gravity \cite{loop}. 

The most important next steps in this research program are, of course, to better understand the nature of the quenched graphs for $d=2$ and, above all, to study quenched graphs at finite temperatures to compare with the antiferromagnetic phase transition derived in mean field theory \cite{tru2}. These points are the subjects of further, ongoing investigation.

\begin{acknowledgments}
I would like to thank F. Biancalana for his active help in the early stages of this work. Many thanks also to O. Knill, Y. Ollivier, G. Perarnau, B. Reed, and P. Romon for explaining the fine points of their mathematical results on graph theory. 
\end{acknowledgments}


\begin{references}
\bibitem{graphrev}For a comprehensive review see: R. Albert and L. Barabasi, {\it Rev. Mod. Phys.} {\bf 74} (2002) 47. 
\bibitem{bia1}For a review see: G. Bianconi and C. Rahmede, {\it Phys. Rev.} {\bf E93} (2016) 032315. 
\bibitem{triang}For a review see: J. Ambjorn, A. G\"orlich, J. Jurkiewicz and R. Loll, {\it Phy. Rep.} {\bf 58} (2012) 127. 
D. P. Rideout and R. D. Sorkin, {\it Phys. Rev. } {\bf D61} (1999) 024002; for a review see: F.Dowker, arXiv:gr-qc/0508109. 
\bibitem{wen} M. Levin and W-G. Wen, {\it Phys. Rev.} {\bf 71} (2004) 045110.
\bibitem{graphity}T. Konopka, F. Markopoulou and L. Smolin, arXiv:hep-th / 0611197; T. Konopka, F. Markopoulou and S. Severini, {\it Phys. Rev.} {\bf D77} (2008) 104029. 
\bibitem{bia2} G. Bianconi, C. Rahmede and Z. Wu, {\it Phys. Rev.} {\bf E92} (2105) 022815; G. Bianconi and C. Rahmede, {\it Scient. Rep} {\bf 5} (2015) 13979. 
\bibitem{sasakura}H. Chen, N. Sasakura and Y. Sato, {\it Phys. Rev.} {\bf D93} (2016) 064071. 
\bibitem{tru1}C. A. Trugenberger, {\it Phys. Rev.} {\bf D92} (2015) 084014.
\bibitem{tru2}C. A. Trugenberger, {\it Phys. Rev.} {\bf E92} (2015) 062818. 
\bibitem{kazakov}V. A. Kazakov, {\it Phys. Lett.} {\bf A119} (1986) 140; D. V. Boulatov and V. A. Kazakov, {\it Phys. Lett.} {\bf B186} (1987) 379; 
\bibitem{gorski}V. Avetisov, A. Gorsky, S. Nechaev and O. Valba, {\it Phys. Rev. } {\bf E93} (2015) 012302. 
\bibitem{hilbert}See e.g.: M. Ahmed and S. Bokhari, {\it IEEE T. Parall. Distrib.} {\bf 18} (2007) 1258. 
\bibitem{rrg}B. Bollobas and W. Fernandez de la Vega, {\it Combinatorica} {\bf 2} (1981) 125;  N. C. Wormald, {\it Surveys in Combinatorics}, London Mathematical Society Lecture Note Series, J. D. Lamb and D. A. Preece eds. Cambridge University Press, Cambridge (1999); S. N. Dorogotsev, {\it Lectures on Complex Networks}, Clarendon Press, Oxford (2010). 
\bibitem{regge}T. Regge, {\it Nuovo Cim.} {\bf 19} (1961) 558; for a review see e.g.: R. M. Williams and P. A. Tuckey, {\it Class. Quant. Gravity} {\bf 9} (1992) 1409. 
\bibitem{quench}For a review see, e.g.: A. J. Bray, {\it Adv. Phys.} {\bf 43} (1994) 357; S. Biswas, arXiv:1603.01646. 
\bibitem{olli1}Y. Ollivier, {\it J. Funct. Anal.} {\bf 256} (2009) 810; Y. Ollivier, {\it Adv. Stud. Pure Math.} {\bf 57} (2010) 343; Y. Linn, L. Lu and S. T. Yau, {\it Tohoku Math. J.} {\bf 63} (2011) 605; B. Loisel and P. Romon, {\it Axioms} {\bf 3} (2014) 119. 
\bibitem{olli2}B. B. Bhattacharya and S. Mukherjee, {\it Discrete Mathematics} {\bf 338} (2015) 23. 
\bibitem{knill} O. Knill, arXiv:1111:5395, arXiv:1205.0306, arXiv:1202.4514, arXiv:1307.3809 arXiv:1501.03116.
\bibitem{loopreg1}B. D. McKay, N. Wormald and B. Wysocka, {\it Elect. J. Combinatorics} {\bf 11}, R66, (2004); H. Garmo, {\it Random Structures \& Algorithms} {\bf 15}, 43, (1999).
\bibitem{loopreg2}F. Joos, G. Perarnau, D. Rautenbach and B. Reed, arXiv1601.03714.
\bibitem{holo}For a review see e.g.: R. Bousso, {\it Rev. Mod. Phys.} {\bf 74} (2002) 825; L Susskind and J. Lindesay, {\it An Introduction to Black Holes, Information and the String Theory Revolution, The Holographic Universe}, World Scientific, Singapore (2005). 
\bibitem{fabio}F. Biancalana, unpublished. 
\bibitem{liu}J. Jost and S. Liu, {\it Discrete Comput. Geom.} {\bf 51} (2014 300. 
\bibitem{geroch} R. P. Geroch, {\it Space-Time Structure from a Global Viewpoint} in {\it General Relativity and Cosmology}, Inter. School of Physics "Enrico Fermi", Academic Press (1971). 
\bibitem{hawking}R. Geroch and G. T. Horowitz, {\it General Relativity; an Einstein Centenary Survey}, S. W. Hawking and W. Israel eds., Cambridge University Press (1979). 
\bibitem{minguzzi}For a comprehensive review see: E. Minguzzi and M. Sanchez, {\it The Causal Hierarchy of Spacetime} in {\it Recent Developments in Pseudo-Riemannian Geometry}, H. Baum and D. Alekseevsky eds., ESI Lect. Math. Phys., 299 (2008). 
\bibitem{spectral} See e.g.: S. N. Dorogovtsev and J. F. F. Mendes, {\it Adv. Phys.} {\bf 51} (2002) 1079.
\bibitem{rwalk}L. Lovasz, {\it Combinatorics} {\bf Paul Erd\"os is Eighty vol. 2} (1993) 1; R. Burioni and D. Cassi, {\it J. Phys. A: Math. Gen.} {\bf 38} (2005) R45, G. Giasemidis, arXiv:1310.8109. 
\bibitem{jonsson}T. Jonsson, {\it Phys. Lett.} {\bf B278} (1992) 89.
\bibitem{maldacena}For a review see: O. Aharony, S. S. Gubser, J. Maldacena, H. Ooguri and Y. Oz, {\it Phys. Rep.} {\bf 323} (2000) 183. 
\bibitem{loop}For a review see: C. Rovelli, {\it Living Rev. Relativity} {\bf 1} (1998) 1; H. Nicolai, K. Peeters and M. Zamaklar, {\it Class. Quant. Grav.} {\bf 22} (2005) R193. 







\end{references}
\end{document}